\documentstyle[11pt,newpasp,epsf]{article}

\begin{document}

\title{Bar-driven Transport of Molecular Gas in Spiral Galaxies: \\
	Observational Evidence}
\author{Kazushi Sakamoto}
\affil{Nobeyama Radio Observatory, Nagano, 384-1305, Japan}

\section*{ }	
The NRO-OVRO CO imaging survey has provided molecular gas distributions
in the centers of 20 nearby spiral galaxies at $\sim 300$ pc
resolution (Sakamoto et al. 1999a). 
It is found from the survey that central condensations of molecular gas 
with sub-kpc sizes and $10^{8}$--$10^{9} M_{\sun}$ masses are prevalent 
in $\sim L^{*}$ galaxies.
Moreover, as shown in Fig. 1, the degree of gas concentration to 
the central kpc 
(estimated from comparison with single-dish data) is 
found to be higher in barred galaxies than in unbarred systems
(Sakamoto et al. 1999b). 
This is the first statistical evidence for the higher central 
concentration of molecular gas (CO) in barred galaxies, 
strongly supporting the theory of bar-driven gas transport. 
To account for the excess gas in barred nuclei, more than half 
of molecular gas in the central kpc of a barred galaxy must have been 
transported there from outside by the bar. 
The time-averaged rate of gas inflow, $\langle \dot{M} \rangle$, 
is statistically estimated 
(through the gas consumption rates estimated from H$\alpha$
and far-IR) to be larger than 0.1 -- 1 $M_{\sun}$ yr$^{-1}$. 
The degree of gas concentration also helps to test
the predictions of bar dissolution and secular
morphological evolution induced by bar-driven gas transport 
(Norman et al 1996; Pfenniger in this volume).
Our current data suggest that bar-dissolution times are longer than the 
consumption times of central gas concentrations in barred galaxies,
and thus prefer slow ($t > 10^{8}$ -- $10^{10}$ yr) bar-dissolution.
A search for non-barred galaxies with high gas concentration, 
presumably galaxies just after quick bar dissolution,
is important to better constrain the bar-dissolution timescale.

\begin{figure*}
{\hfill\epsfysize=8cm\epsfbox{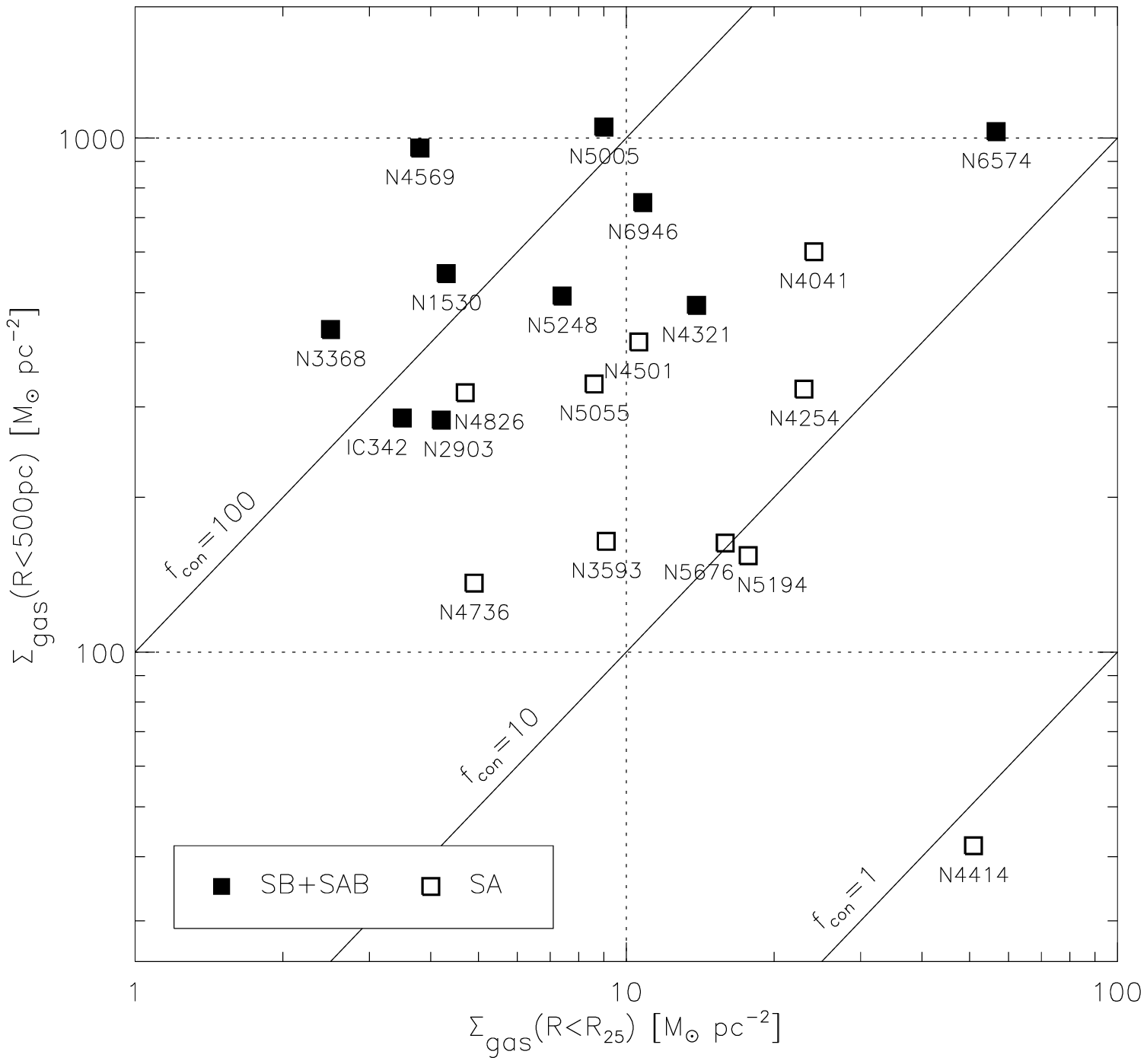}	
 \hspace{0.5cm}		
 \epsfysize=8cm\epsfbox{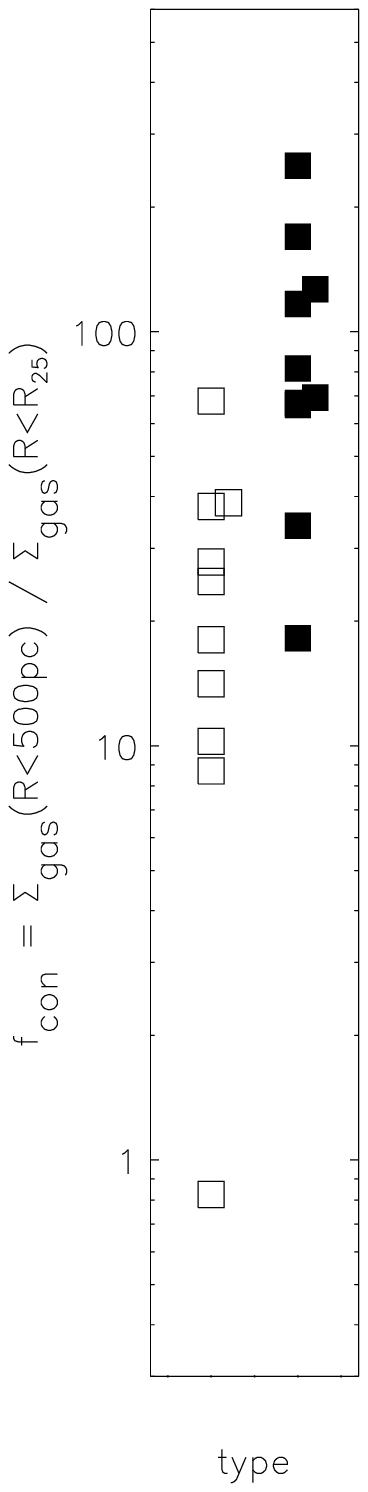}	
\hfill}
\caption{Concentration factor 
$f_{\rm con} \equiv \Sigma_{\rm gas}\left( R \leq {\rm 500 pc} \right) /
		    \Sigma_{\rm gas}( R \leq R_{25} )$
measures degree of gas concentration in galaxies.
Barred galaxies have higher $f_{\rm con}$ than unbarred counterparts 
(Sakamoto et al. 1999b).}
\end{figure*}

There have been a few other lines of observational evidence for the bar-driven
gas transport.  
\begin{table}
\caption{Observational evidence for bar-driven gas transport}
\begin{tabular}{cccccc}
\tableline 
\tableline
Method			& obs.	 	& evidence 		& stat. 	& $\dot{M}$ 			& ref. \\
			&		& Barred gals have ...  &		&  &   \\
\tableline
gas concentration	& CO 		& higher $f_{\rm con}$ 	& yes 		& $\langle \dot{M} \rangle$ 	& (1) \\
central SFR 		& H$\alpha$ etc. & higher SFR		& yes		& no				& (2) \\
metallicity gradient	& opt.		& shallower gradients	& yes		& $\langle \dot{M} \rangle$	& (3) \\
dynamical modeling	& CO, NIR	& net inward gas flow	& no (yet)	& $\dot{M}$			& (4) \\
obscuration of AGN	& X-ray		& larger absorption	& yes		& no				& (5) \\
\tableline
\end{tabular}
References: 
(1) Sakamoto et al. 1999b; 
(2) Ho et al. 1997 and references therein;
(3) Roy 1996 for a review;
(4) Quillen et al. 1995, Regan et al. 1997;
(5) Maiolino et al. 1999.
\end{table}
Table 1 summarizes the pieces of evidence and their properties.
All of them support bar-driven gas inflow and, though each of them provides
different types of information, they are complementary with each other 
(e.g., some of them are statistical and the others are not, 
some can give $\langle \dot{M} \rangle$ while others give instantaneous $\dot{M}$).
A next step would be to combine these methods, with increasing
sample size, in order to construct a sequence of mass transfer,
star formation, and morphological changes in galaxies.
Galaxy evolution along the Hubble sequence
is now within the reach of observational tests.

\acknowledgments
The CO survey and the subsequent analysis were made 
in collaboration with S. K. Okumura, 
S. Ishizuki, and N. Z. Scoville.

\end{document}